\documentclass[epj,nopacs]{svjour}
\usepackage{Ratcliffe}
\usepackage[dvips]{hyperref}
\hypersetup{%
  pdfauthor={Philip G. Ratcliffe <philip.ratcliffe@uninsubria.it>},
  pdftitle={Transversity K Factors for Drell-Yan Processes},
  pdfsubject={submitted to the European Physical Journal, section C},
  pdfkeywords={transversity, transverse spin, parton model},
}
\begin{document}
%——————————————————————————————————————————————————————————————————————————————%
\begin{fmffile}{fmf}
%——————————————————————————————————————————————————————————————————————————————%
\title{Transversity \maybebm{K} Factors for Drell--Yan Processes}
\author{Philip G. Ratcliffe\inst{1,2}}
\institute{%
  Dipartimento di Fisica e Matematica, Universit\`{a} degli Studi
  dell'Insubria---sede di Como\\
  via Valleggio~11, 22100 Como, Italy
\and
  Istituto Nazionale di Fisica Nucleare---sezione di Milano\\
  via G.~Celoria~16, 20133 Milano, Italy\\
  \textsf{philip.ratcliffe@uninsubria.it}
}
\date{Received: date / Revised version: date}
\abstract{%
  The question of the $K$ factor in transversely polarised Drell--Yan (DY)
  processes is examined. The transverse-spin case is peculiar for the absence
  of a reference point in deeply inelastic scattering (DIS). Therefore, in
  order to study more fully the possible effects of higher-order corrections on
  DY asymmetries, a DIS definition for transversity is devised using a
  hypothetical scalar (Higgs-like) vertex. The results show that some care may
  be required in interpreting experimentally extracted partonic transversity,
  in particular when comparing with model calculations or predictions.
}
%——————————————————————————————————————————————————————————————————————————————%
\maketitle
%——————————————————————————————————————————————————————————————————————————————%
\section{Introduction}
The theoretical framework for describing transversity (at the basic level of
partonic processes, \ac{QCD} evolution, radiative effects \etc) is now solid
\cite{Barone:2001sp} and a number of experiments aimed at its measurement are
on-line or under development: HERMES \cite{Rostomyan:2003zr}, COMPASS
\cite{LeGoff:2002qn} and the RHIC spin programme \cite{Bland:2004um}; there are
also proposals for \ac{DY} measurements with polarised antiprotons in the High
Energy Storage Ring at GSI \cite{Bertini:2004l1, Rathman:2004l1} (related
preliminary theoretical studies have been made regarding access to transversity
in $J/\psi$ production~\cite{Anselmino:2004ki, Barone:2005a1}).

Transversity is the last remaining piece in the partonic jig-saw puzzle
composing the hadronic picture. However, the standard procedure of adopting
\ac{DIS} as the process to \emph{define} parton densities at the \ac{NLO}
cannot be extended to transversity in a simple manner since it does not
contribute to \ac{DIS}. Furthermore, transverse-spin effects are notoriously
surprising; \eg, see the large and (historically) unexpected \acp{SSA}
\cite{Adams:1991cs, Bravar:1996ki}. Such considerations render imperative the
complete understanding of \ac{NLO} corrections in \ac{DY} before attempts are
made to extract the partonic transversity distributions. See, \eg,
Ref.~\cite{Barone:2005au} for a detailed discussion of transversity and also
\acp{SSA}.

One might instead consider double-spin asymmetries $A_{TT}$ for other
processes, such as: $p^{\uparrow}p^{\uparrow}\to\text{jet}+X$, $\gamma+X$ \etc.
Unfortunately, however, predictions for $A_{TT}$ always turn out to be very
small \cite{Soffer:2002tf, Mukherjee:2003pf}, so that measuring transversity
directly appears feasible only in doubly polarised $p\bar{p}$ interactions.

Since all \ac{QCD} and \ac{EW} vertices conserve quark chirality, transversity
actually decouples from \ac{DIS}. Chirality flip is not a problem though if the
quark lines connect to different hadrons as in, \eg, the \ac{DY} process.
Unfortunately, there is a \caveat to accessing transversity in DY: Hikasa's
theorem \cite{Hikasa:1986qi}, which states that, owing to chiral symmetry,
transversity effects vanish upon integrating over the lepton-pair azimuth. No
simple proof of the theorem exists (it has to do with the $\gamma$-matrix
algebra). Let us now make a few observations based on these properties of
transversity:
\begin{enumerate}
\item
the only ``gold-plated'' process in which transversity may be measured directly
(\ie, without the need of more-or-less exotic fragmentation functions) is
\ac{DY};
\item
Hikasa's theorem implies the use of a slightly less than fully inclusive
process, in as much as one angle must be left unintegrated;
\item
in the case of transversity asymmetries, helicity conservation may not
necessarily provide the usual safeguard against large $K$ factors.
\end{enumerate}
The above have non-trivial implications with respect to the measurement of
transversity in \ac{DY} and interpretation of the results.

\paragraph{1.}
In the absence of a \ac{DIS} reference point, there is no immediate way of
fully evaluating the possible importance of higher-order \ac{QCD} corrections.
The $K$ factors are known to be large at the level of cross-sections in both
the unpolarised \cite{Altarelli:1979ub} and helicity-dependent
\cite{Ratcliffe:1983yj} cases. However, in the helicity case the large
corrections cancel in the asymmetry \cite{Ratcliffe:1983yj}. To a large extent
this cancellation can be traced to the conservation of helicity along fermion
lines in gauge theories---the $\Order{\alpha_s}$ Wilson coefficient for the
\ac{DY} process is identical for the helicity-dependent and -independent
pieces, as too are the \ac{LO} anomalous dimensions. Note that in the case of
heavy-flavour production the corrections to the helicity asymmetry are large,
precisely because mass terms introduce helicity flip, destroying the usual
protection.

\paragraph{2\,+\,3.}
The coefficient function for transversely polarised \ac{DY} differs
significantly from the other two cases \cite{Vogelsang:1993jn,
Vogelsang:1998ak}. Moreover, the \ac{LO} anomalous dimensions differ---there is
no corresponding conserved quantity or sum-rule.

Given the marked differences from the other two cases, it may be useful to
examine the question of \ac{DY} $K$ factors for transversity. In order to do
this, it is clearly necessary to find some suitable \ac{DIS}-like process as a
reference point. The principal requirement is a spin-flip mechanism. There are
two obvious possibilities \apriori: either a quark mass term or a scalar
vertex. Now, of course, \ac{DIS} with transversely polarised leptons and
nucleons should be considered and therefore the twist-three structure function
$g_2$ for general reviews) is the natural object of study (see \eg,
Refs.~\cite{Jaffe:1990xx, Barone:2003fy}. It turns out, however, that although
transversity is intimately related to the evolution of $g_2$ (the relevant
operator is indeed proportional to the quark mass \cite{Antoniadis:1981dg,
Bukhvostov:1985rn, Ratcliffe:1986mp}), at the level of direct contribution to
polarised \ac{DIS} it actually cancels against other higher-twist contributions
owing to the equations of motion, see for example Ref.~\cite{Anselmino:1995gn}.
Although the calculation is rather delicate, the possibility of defining a
coefficient function for transversity via its \role in the evolution of $g_2$
has been examined \cite{Rossi:2002t1}, with similar results to those presented
here.

A simpler and more direct approach is to identify a \ac{DIS}-like process in
which a scalar particle plays a \role. Since the Higgs boson does indeed
interact with quarks (as with leptons too), the obvious solution to the problem
is a \gedanken process in which the exchange is no longer via the electroweak
gauge fields but via the Higgs particle. To be precise, in order to obtain the
required single spin-flip, Higgs--Vector interference diagrams actually need to
be considered. Of course, there is no intended suggestion here that such a
process should really be measured, but merely that it forms a suitable basis
for a theoretical cross-check. We should remark that such a process has
effectively already been exploited for the calculation of $h_1(x)$ itself
\cite{Ioffe:1995aa}, on the basis of a suggestion by Jaffe. In any case,
various tests will be performed to ensure that the results do not depend on the
specific nature of the vertex introduced.

Before moving on to the calculation, let us spend a few more words on the
physical significance of the $K$-factor. While at a theoretical level the
meaning of higher-order corrections to any given process is clear, at a
phenomenological level in the parton-model there is an inherent ambiguity owing
to the necessary input of the parton densities themselves. Indeed, the
$K$-factor was used historically to represent the discrepancy between
experimental results for the \ac{DY} cross-section and the \ac{LO} theoretical
predictions based on parton densities extracted from \ac{DIS} and thus \emph{by
definition} (as in Refs.~\cite{Altarelli:1979ub, Ratcliffe:1983yj}) the
phenomenological $K$-factor is the translation factor from \ac{DIS} to \ac{DY}.
Therefore, since all model calculations or estimates of partonic transversity
densities rely to some extent on \ac{DIS} for overall normalisation or
determination of model parameters, self-consistency would require a procedure
of the type to be described here. Since, furthermore, the overall $K$-factor
so-defined receives large contributions from \emph{both} \ac{DIS} \emph{and}
\ac{DY}, this is a non-trivial observation. The peculiar structure of
transversity leaves room for very different corrections as compared to the
spin-averaged or helicity-dependent cases, for both \ac{DIS} and \ac{DY}
independently.

%Of course, an alternative approach might be to work entirely in a \ac{DY}-based
%scheme (\ie, for both the unpolarised and polarised analyses); this would,
%however, necessitate some reworking of most model predictions.

Thus, in the following section the calculations are described, the
Higgs--Vector interference mechanism is examined in detail and \ac{NLO}
calculation of the related Wilson coefficients is performed. The known results
for the \ac{DY} process are discussed and finally the relevant $K$ factors are
extracted. In the closing section some conclusions are drawn and comments
relevant to future measurements of transversity via \ac{DY} scattering are
made.%
\footnote{%
  Owing to correction of an error in the code used for numerical estimates,
  the results shown here are a little less dramatic than those presented by the
  author in past conferences.
}%

\section{The Calculation}

\subsection{Drell--Yan cross-section and asymmetries}

It is now standard to define the helicity- and transversity-weighted
cross-sections by
\begin{subequations}
\begin{align}
  \frac{\D\DL\sigma}{\D{Q^2}} &\equiv
  \frac12
  \Bigg[
    \frac{\D\sigma^{++}}{\D{Q^2}} -
    \frac{\D\sigma^{+-}}{\D{Q^2}}
  \Bigg]
\shortintertext{and}
  \frac{\D\DT\sigma}{\D{Q^2}} &\equiv
  \frac12
  \Bigg[
    \frac{\D\sigma^{\,\uparrow\uparrow}\;}{\D{Q^2}} -
    \frac{\D\sigma^{\,\uparrow\downarrow}\;}{\D{Q^2}}
  \Bigg]
  ,
\end{align}
\end{subequations}
where the prefixes $\DL$ and $\DT$ indicate longitudinal-spin (or helicity) and
transverse-spin (or transversity) dependence respectively, $\pm$ refer to
initial-state proton helicities and $\uparrow,\downarrow$ to transverse
polarisations. The double-spin asymmetries are then
\begin{subequations}
\begin{align}
  A_{LL} &\equiv \frac{\D\DL\sigma/\D{Q^2}}{\D\sigma/\D{Q^2}}
\shortintertext{and}
  A_{TT} &\equiv \frac{\D\DT\sigma/\D{Q^2}}{\D\sigma/\D{Q^2}}
  .
\end{align}
\end{subequations}
The large \ac{NLO} corrections afflict both the numerators and denominators.
The question is to what extent they are correlated, \ie, to what extent they
are the same and thus cancel in the ratio.

Turning then to the calculation of the $K$ factor, the procedure will be
essentially identical to that followed in earlier work \cite{Altarelli:1979ub,
Ratcliffe:1983yj} and thus we shall not dwell on the general technicalities,
save for those points that are significantly different in the case of
transverse polarisation. The first peculiar aspect to be exploited is that,
owing to the charge-conjugation properties of the relevant operator, the
evolution of transversity is of the flavour \ac[\emph]{NS} type. In the \ac{NS}
case the effect of higher-order corrections may be represented in the following
schematic way:
\begin{multline}
  F(x,t) =
  \int_x^1 \frac{\D{y}}y
  \sum_f Q_f^2
  \bigg[
    \delta\!\Big(1-\frac{x}{y}\Big)
\\
  \null
    +
    \frac{\alpha_s(Q^2)}{2\pi} \, t \, P\!\Big(\frac{x}{y}\Big)
    +
    \frac{\alpha_s(Q^2)}{2\pi} \, C\!\Big(\frac{x}{y}\Big)
  \bigg]
  q_f(y,t) ,
  \label{eq:parton-formula}
\end{multline}
where $t\equiv\ln(Q^2/\mu^2)$, with $Q^2$ the virtuality of the photon, $q_f(y,t)$
and $Q_f$ are respectively the parton density and charge of quark flavour $f$,
$P$ is the universal quark--quark splitting function and $C$ the
process-dependent Wilson coefficient. The quantity $F(x,t)$ on the \ac{LHS}
then represents a generic (flavour \ac{NS}) structure function and the three
terms inside the square brackets on the \ac{RHS} represent: 1.~the \ac{LO}
point-like contribution; 2.~the \ac{LL} correction; and 3.~the \ac{NLO}
correction. It is this last that is of interest here.

To \ac{NLO} the \ac{DY} cross-section for $p\bar{p}$ scattering is expressed in
term of parton densities as follows:
\begin{align}
  Q^2 \frac{\D\sigma^\text{DY}}{\D{Q^2}} &=
  \frac{4 \pi \alpha}{9 s}
  \int_0^1 \D{x_1} \D{x_2} \D{z} \; \delta (x_1x_2z-\tau)
  \nonumber
\\
  & \hspace{1em} \null \times
  \sum_f Q_f^2
  \bigg[ q_f(x_1,Q^2) \, \bar{q}_f(x_2,Q^2) + (1 \leftrightarrow 2) \bigg]
  \nonumber
\\
  & \hspace{4em} \null \times
  \bigg[ \delta (1-z) + \frac{\alpha_s(Q^2)}{2 \pi} \, C^\text{DY}\!(z) \bigg]
  ,
  \label{eq:DY-unpol}
\end{align}
where $\tau=Q^2/s$, $Q^2$ is the invariant mass squared of the lepton pair and
$s$ is the total hadron \ac{CM} energy squared. In Eq.~\eqref{eq:DY-unpol}
$x_{1,2}$ are the momentum fractions carried by the (anti)quarks inside hadrons
$1$ and $2$ respectively. It is then the difference between the \ac{DIS}
corrections, with which the \ac{NLO} parton distributions are defined via
Eq.~\eqref{eq:parton-formula}, and \ac{NLO} \ac{DY} corrections that
constitutes the phenomenological $K$ factor.

The \ac{LL} splitting functions $P$ are well known \cite{Gribov:1972ri,
Lipatov:1975qm, Altarelli:1977zs, Dokshitzer:1977sg, Baldracchini:1981uq,
Artru:1990zv} and may be expressed in the following compact form:
\begin{subequations}
\begin{align}
  \DL{P}(z)
  &=
  P(z)
  =
  \CF \left[ \frac{1+z^2}{1-z} \right]_+
\\[-3ex]
\intertext{and}
\nonumber\\[-5ex]
  \DT{P}(z)
  &=
  P(z) - \CF (1-z) ,
\end{align}
\end{subequations}
The definition of the so-called ``plus'' regularisation is recalled in
Appendix~\ref{sec:Plus-reg}. Already then it is evident that although
fermion-helicity conservation guarantees identical evolution for \ac{NS}
spin-averaged and helicity-weighted quark densities, the same does not hold for
the transversity case.

The problem now is to calculate the coefficient $C(z)$ of the third term in
Eq.~\eqref{eq:parton-formula}. This must be done for both the \ac{DIS} and
\ac{DY} processes. As is well-known, a large part of the \ac{DY} $K$ factor can
be attributed to the change from a space-like $Q^2$ in \ac{DIS} to time-like in
\ac{DY}. However, this is not the only origin of large corrections and one
should be concerned that the transversity case, with the extra requirement on
the final-state phase space in \ac{DY}, might introduce important differences.

\subsection{The Drell--Yan Process}

Since the results for the \ac{DY} process are known \cite{Vogelsang:1993jn,
Vogelsang:1998ak}, it is perhaps better to begin with this coefficient.
%The partonic subprocess to be calculated is shown in
%Fig.~\ref{fig:qq-gp-scattering}.
%\begin{figure}
%  \centering
%  \input{texfiles/qq-gp-scattering}
%  \caption{%
%    The two diagrams contributing to the \ac{NLO} \ac{DY} hard partonic
%    $q\bar{q}\to\gamma^*g$ scattering subprocess.
%  }
%  \label{fig:qq-gp-scattering}
%\end{figure}
The virtual photon decays into a final lepton pair, of which then the azimuthal
angle must be left unintegrated in the case of transversity. In the \ac{MMS}
scheme the results for the unpolarised \cite{Altarelli:1979ub}, helicity
\cite{Ratcliffe:1983yj} and transversity \cite{Vogelsang:1998ak} \ac{DY}
coefficient
functions are as follows:%
\footnote{%
  Here and in what follows only the unpolarised result will be presented in
  full while the helicity and transversity cases will be shown as differences
  with respect to the unpolarised case.
}%
\begin{subequations}
\begin{align}
  C^\text{DY}\!(z) &=
  C_L^\text{DY}\!(z)
  \nonumber
\\[1ex]
  & =
  \CF
  \bigg\{
    4(1+z^2) \left[ \frac{\ln(1-z)}{1-z} \right]_+
    - \frac{2(1+z^2)\ln z}{(1-z)}
  \nonumber
\\[1ex]
  & \hspace{8em} \null
    + \left[ \frac{2}{3} \pi^2 - 8 \right] \delta(1-z)
  \bigg\} ,
\\[1ex]
  C_T^\text{DY}\!(z) &= C^\text{DY}\!(z)
  + \CF
  \bigg\{
    -4(1-z) \ln(1-z)
  \nonumber
\\[1ex]
  & \hspace{1.7em} \null
    + 2(1-z)\ln z
    - \frac{6z\ln^2z}{(1-z)} + 4(1-z)
  \bigg\} .
\end{align}
\end{subequations}
It is important to note that although the large $\pi^2$ terms and indeed the
coefficient of the $\delta$-function (which indeed constitute the bulk of the
$K$ factor) appear invariant here, in the transversity case there is a new
term, $-\tfrac{6z\ln^2z}{(1-z)}$, not found in the others.

\subsection{Deeply Inelastic Scattering}

\begin{figure}
  \centering
  \input{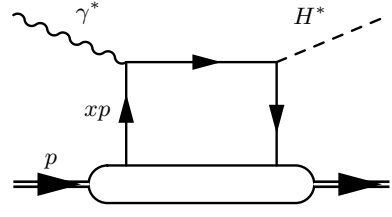}
  \caption{%
    The \ac{DIS} ``handbag'' diagram for a photon--Higgs interference process.
  }
  \label{fig:Higgs-photon}
\end{figure}

In order to accommodate spin-flip in the standard \ac{DIS} ``handbag'' diagram,
one of the vertices should involve a Higgs-like scalar, see
Fig.~\ref{fig:Higgs-photon}.
The contribution of this diagram can be expressed in terms of a structure that
will be called $h_1$ here for brevity. Projecting with
$\gamma_5\slashed{p}\slashed{s}_T$ then leads to
\begin{equation}
  W^\mu = h_1(x,Q^2) \, \frac{i \epsilon^{q p s_T \mu}}{p{\cdot}q} .
\end{equation}

\subsubsection{Real-gluon contributions}

\begin{figure}
  \centering
  \input{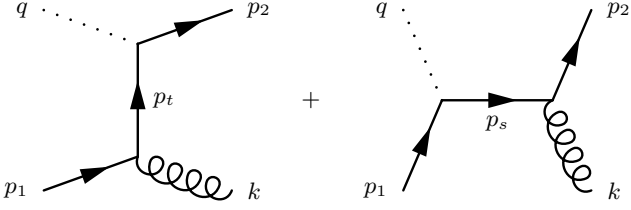}
  \caption{%
    The two diagram types contributing to the \ac{NLO} \ac{DIS} hard partonic
    $q\gamma^*(H^*)\to{}gq$ scattering subprocess, the dotted line represents
    either a virtual photon or Higgs.
  }
  \label{fig:qp-gq-scattering}
\end{figure}

The \ac{NLO} Wilson coefficient may now be calculated from the diagrams in
Fig.~\ref{fig:qp-gq-scattering}.
The process to be calculated is, of course, still photon--Higgs interference.
The use of dimensional regularisation poses the problem of dealing with
$\gamma_5$, which naturally arises in the case of polarised \ac{DIS} (for both
helicity and transversity) owing to the projector
$\gamma_5\slashed{p}\slashed{s}_T$. Note that for the $q\bar{q}\to\gamma^*g$
\ac{DY} subprocess this is not a problem since both the quark and antiquark
bring one power of $\gamma_5$, which then cancels before calculating any
traces. The technique adopted here is that of defining a fully anti-commuting
but non-cyclic $\gamma_5$, see for example Ref.~\cite{Korner:1992sx}; see also
Ref.~\cite{Jegerlehner:2000dz} for a detailed discussion of this technique.
Consistency then requires that all traces be evaluated from the same reference
point, which in the usual \ac{DIS} case is unambiguously one of the photon
vertices. Here the scalar vertex could also be chosen---an explicit check shows
that there is no ambiguity in the results.

In the \ac{MMS} scheme (see Appendix~\ref{sec:Msbar} for a working definition),
adopting the above-mentioned $\gamma_5$ scheme and suppressing (for clarity) a
common factor
\begin{equation*}
  \CF \, \frac{\Gamma^2(1-\epsilon)}{\Gamma(1-2\epsilon)}
\end{equation*}
on the \ac{RHS} of all equations, the results for the real contributions of the
diagrams in Fig.~\ref{fig:qp-gq-scattering} are:
\begin{subequations}
\begin{align}
  \widetilde{C}^\text{DIS-R}(z) &=
  \frac2{\epsilon^2} \, \delta(1-z)
  - \frac1{\epsilon} \left[ \frac{1+z^2}{\,(1-z)_+} - \frac32\delta(1-z) \right]
  \nonumber
\\
  & \hspace{-1em} \null
  + (1+z^2) \left[\frac{\ln(1-z)}{1-z}\right]_+
  - \frac32 \frac1{\,(1-z)_+}
  + 3 + 2z
  \nonumber
\\
  & \hspace{3em} \null
  - (1+z^2) \frac{\ln z}{(1-z)}
  + \frac72 \, \delta(1-z) ,
\\
  \widetilde{C}_L^\text{DIS-R}(z) &=
  \widetilde{C}^\text{DIS-R}(z) - 1 - z ,
\\
  \widetilde{C}_T^\text{DIS-R}(z) &=
  \widetilde{C}^\text{DIS-R}(z)
  + \frac1{\epsilon} (1-z)
  \nonumber
\\
  & \hspace{3em} \null
  - (1-z)\ln\bigg(\frac{1-z}{z}\bigg)
  - \frac32 - 2z
  ,
  \label{eq:combined-trans}
\end{align}
\end{subequations}
where the $\widetilde{C}^\text{DIS-R}(z)$ are defined to be the combined
quantities
\begin{equation}
  \widetilde{C}^\text{DIS}(z) =
  t \, P(z) + C^\text{DIS}(z),
\end{equation}
with $P(z)$ and $C(z)$ being replaced respectively by $\DL{P}(z)$ and $C_L(z)$
\etc, where necessary. In the first equation, for $\widetilde{C}(z)$, an
additional contribution $3z$ due to $F_L$ has been included to give the
correction corresponding to the use of $F_2$ to define $q(x)$
\cite{Altarelli:1979ub}. Moreover, in Eq.~\eqref{eq:combined-trans} the
remaining $\epsilon$ is due to the difference in splitting functions and
disappears in the final expression for the full coefficient.

To extract the desired coefficients $C^\text{DIS-R}(z)$, the virtual
corrections must now, of course, also be added. First however, note that the
results for the unpolarised and helicity cases agree with previous
calculations, \cite{Altarelli:1979ub} and \cite{Ratcliffe:1983yj} respectively.
Note also that the results for the various cases are (not surprisingly)
similar: while the coefficient for $h_1$ is a little different (owing to the
finite residues of the \ac{UV} divergences, which lead to different splitting
functions), the \ac{IR} double poles in $\epsilon$ are identical (and in any
case cancel with the virtual contributions) and the single poles themselves
are, of course, to be absorbed into the scale-dependent parton densities. In
particular, there is no trace of the $\tfrac{6z\ln^2z}{(1-z)}$ term found in the
\ac{DY} coefficient.

\subsubsection{Virtual-gluon contributions}

The virtual contributions can be partially gleaned from the literature;
however, the scalar-vertex correction remains to be evaluated and this requires
a some care. The real and virtual contributions are separately gauge invariant
and a natural choice (as in Refs.~\cite{Altarelli:1979ub, Ratcliffe:1983yj} and
other cited work) is the Landau gauge, where it is only the vertex correction
that need be calculated. Schematically,
\begin{subequations}
\begin{align}
  \Gamma_V^\mu(q^2) &= \gamma^\mu
  \left[1 + \frac{\alpha_s}{2\pi} \, \delta_V\right]
\shortintertext{and}
  \Gamma_S(q^2) &= \mskip5mu \UnitOp \mskip4mu
  \left[1 + \frac{\alpha_s}{2\pi} \, \delta_S\mskip1mu\right]
  ,
\end{align}
\end{subequations}
where $\UnitOp$ represents the ``bare'' scalar vertex. In \ac{MMS} the following
results are then obtained:
\begin{subequations}
\begin{align}
  \delta_V &=
  \CF \left[\frac{\mu^2}{-q^2}\right]^\epsilon
  \frac{\Gamma^2(1-\epsilon)}{\Gamma(1-2\epsilon)}
  \left[ - \frac2{\epsilon^2} -\frac3\epsilon - 8 - \tfrac23\pi^2 \right]
\shortintertext[0.5]{and}
  \delta_S &=
  \CF \left[\frac{\mu^2}{-q^2}\right]^\epsilon
  \frac{\Gamma^2(1-\epsilon)}{\Gamma(1-2\epsilon)}
  \left[ - \frac2{\epsilon^2} - 2 - \tfrac23\pi^2 \right]
  .
\end{align}
\end{subequations}
Noting that these corrections multiply $\delta(1-z)$, one immediately sees that
the double poles $1/\epsilon^2$, of \ac{IR} origin, cancel against the real
diagrams, just as they should. However, note also that the single-pole
structure is manifestly different; let us examine this a little more closely.

There is a substantial difference between a vector and a scalar vertex: the
former is related to a conserved current, the latter not. Thus, the vector
current is not renormalised while the scalar does receive radiative
corrections. In other words, one must also take into account the
renormalisation of the coupling constant (\ie, the quark mass in the true Higgs
case) associated with the vertices in consideration.%
\footnote{%
  This observation was made by Bl\"{u}mlein \cite{Blumlein:2001ca} in regard of
  similar calculations aimed at evaluating the evolution kernel.
} %
Indeed, the simplest way to evaluate the contribution is to calculate the
quark-mass renormalisation, including the constant pieces. In the \ac{MMS}
scheme the standard calculation gives
\begin{equation}
  \delta_m =
  -\CF \, \frac{\alpha_s(Q^2)}{2\pi}
  \left[\frac{\mu^2}{-q^2}\right]^\epsilon
  \frac{\Gamma^2(1-\epsilon)}{\Gamma(1-2\epsilon)}
  \frac3\epsilon \, \frac1{1-2\epsilon}
  .
\end{equation}
Including this as a contribution to the virtual corrections, one finally
obtains
\begin{equation}
  \delta_S^\text{full} = \delta_V .
\end{equation}
Thus, combining real and virtual contributions, the complete set of
coefficients for \ac{DIS} are
\begin{subequations}
\begin{align}
  C^\text{DIS}(z) &=
  \CF \bigg\{
  (1+z^2)\left[\frac{\ln(1-z)}{1-z}\right]_+
  - \frac32\frac1{(1-z)_+}
  \nonumber
\\
  & \hspace{4.5em} \null
  + 3 + 2z
  - (1+z^2)\frac{\ln z}{(1-z)}
  \nonumber
\\
  & \hspace{7em} \null
  - \left[ \frac92 + \frac{\pi^2}3 \right] \delta(1-z) \bigg\} ,
\\[1ex]
  C_L^\text{DIS}(z) &=
  C^\text{DIS}(z)
  -1 - z ,
\\[1ex]
  C_T^\text{DIS}(z) &=
  C^\text{DIS}(z)
  \nonumber
\\
  & \hspace{3em} \null
  - \frac32 - 2z
  - (1-z) \ln\left(\frac{1-z}{z}\right) ,
\end{align}
\end{subequations}

\subsection{The \maybebm{K}-Factor Results}

The \ac{DY} and \ac{DIS} coefficients can now be combined to provide a
theoretical $K$ factor. Note that the \ac{DIS} coefficient appears with a
factor two in the required difference; this merely reflects the two quarks (or
rather quark--antiquark) in the initial state for \ac{DY}. The final results
are%
\footnote{%
  A brief review of the results presented here may be found in a recent
  contributed talk~\cite{Ratcliffe:2004wg}.
}%
\begin{subequations}
\begin{align}
  C^\text{DY}\!(z) - 2
  C^\text{DIS}(z)
  &= \CF
  \bigg\{
    2\left(1+z^2\right) \left[\frac{\ln(1-z)}{1-z}\right]_+
  \nonumber
\\
  & \hspace{-6.5em} \null
    + \frac3{(1-z)_+}
    - 6 - 4z
    + \left[\frac43\pi^2+1\right]\delta(1-z)
  \bigg\} ,
\\[1ex]
  C_L^\text{DY}\!(z) - 2
  C_L^\text{DIS}(z)
  &=
  C^\text{DY}\!(z) - 2
  C^\text{DIS}(z)
  \nonumber
\\
  & \hspace{6.7em} \null
  + \CF \, 2(1+z) ,
\\[1ex]
  C_T^\text{DY}\!(z) - 2
  C_T^\text{DIS}(z)
  &=
  C^\text{DY}\!(z) - 2
  C^\text{DIS}(z)
  \nonumber
\\
  & \hspace{-5em} \null
  + \CF
  \bigg\{
    7 - \frac{6z\ln^2z}{(1-z)} - 2(1-z) \ln(1-z)
  \bigg\}
  ,
\end{align}
\end{subequations}
where the origins of the large differences in the last line may thus be traced
in part to the different phase-space restrictions in the transversity case and
in part to the residues due to the different splitting functions. It is perhaps
worth reminding the reader that the bulk of the large $K$ factor in the
unpolarised and helicity cases (the $\pi^2$ terms) comes from the necessary
continuation of $Q^2$ from space-like (in \ac{DIS}) to time-like (in \ac{DY}).
However, the transversity correction contains other non-negligible pieces.

\begin{figure}
  \centering
  \psfrag{DY-2*DIS coefficient difference}
         {\hspace*{3em}$C_i^{\text{DY}(n)}-2C_i^{\text{DIS}(n)}$}
  \psfrag{n}{\hspace*{7em}$n$}
  \psfrag{T1}[Br]{\small spin averaged\hspace*{-1em}}
  \psfrag{T2}[Br]{\small helicity weighted\hspace*{-1em}}
  \psfrag{T3}[Br]{\small transversity weighted\hspace*{-1em}}
  \includegraphics[width=0.45\textwidth]{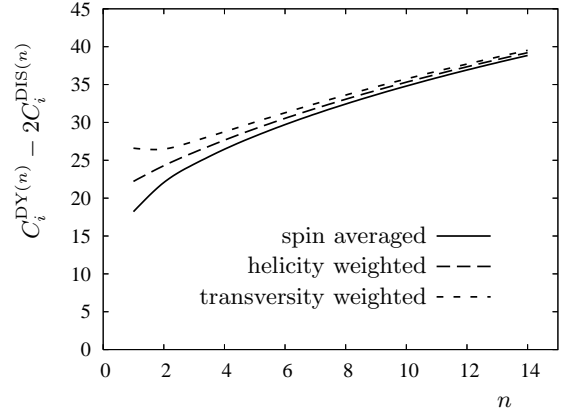}
  \caption{%
    The \acs{DY}--\acs{DIS} difference in the Mellin moments of the Wilson
    coefficients (\ie, the $K$ factor) for the three leading-twist densities.
  }
  \label{fig:coefdiff}
\end{figure}

For a first visual comparison, Fig.~\ref{fig:coefdiff} shows the Mellin
moments, defined by
\begin{equation}
  f^{(n)} \equiv \int_0^1 \D{x} \, x^n \, f(x) ,
\end{equation}
of the above differences in the Wilson coefficients between \ac{DY} and
\ac{DIS} for the three leading-twist densities.
While it is clear that the difference is generally only a little larger for
transversity, the rapidly growing difference as $n\to0$ (equivalent in $z$
space to $z\to0$) with respect to both the unpolarised \emph{and} helicity
cases is particularly striking. Note also that the small-$z$ (small-$n$) region
counts heavily in the convolution integrals, therefore the corrections to the
transversity-weighted cross-section can be significantly larger.

\begin{figure*}
  \centering
  \psfrag{tau}{$\tau$}
  \psfrag{A_LL}[Bl]{$A_{LL}$}
  \psfrag{A_TT}[Bl]{$A_{TT}$}
  \psfrag{LO}[Br]{\small LO}
  \psfrag{NLO}[Br]{\small NLO}
  \includegraphics[width=0.45\textwidth]{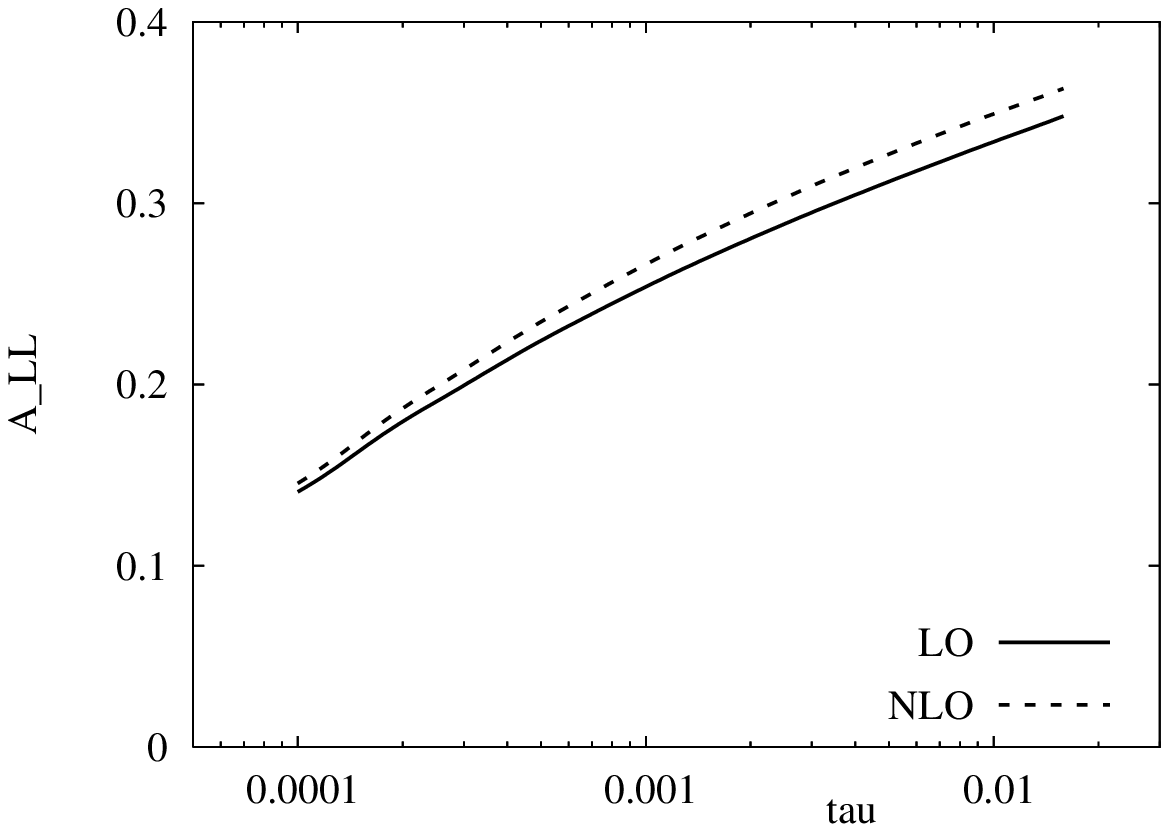}
  \includegraphics[width=0.45\textwidth]{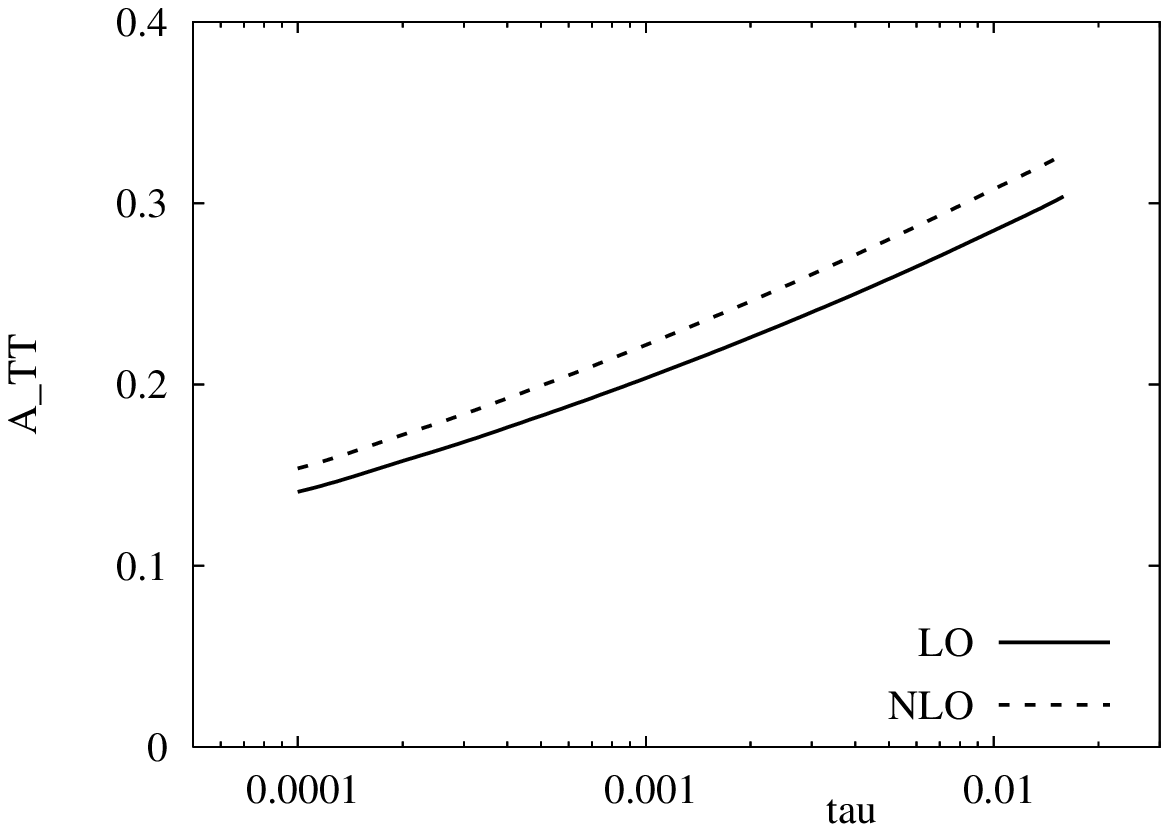}
  \raisebox{2ex}[4ex][0pt]{\hspace{10mm}(a)\hspace{0.43\textwidth}(b)}
  \caption{%
    The doubly polarised $p\bar{p}$ helicity and transversity asymmetries
    $A_{LL}$ and $A_{TT}$ for purely valence-driven \acs{DY} at \ac{LO} and
    \ac{NLO} as functions of $\tau=Q^2/s$, for $\sqrt{s}=200$\,GeV.
  }
  \label{fig:dyasym}
\end{figure*}

To provide an idea of the effect these corrections might have on an
experimentally measured \ac{DY} asymmetry, in Fig.~\ref{fig:dyasym} the
helicity and transversity asymmetries for purely \ac{NS} contributions (in both
the numerator and denominator) are plotted as functions of $\tau$. The
asymmetries are shown for a centre-of-mass energy $\sqrt{s}=200$\,GeV,
corresponding to recent polarised RHIC energies \cite{Bland:2004um}. For the
transversity distribution we have taken a conservative starting point of
$\DT{q}(x,Q_0^2)=\DL{q}(x,Q_0^2)$. The evolution of the distributions has then
been performed to \ac{LO} here as the coefficient differences are scheme
independent and the effect of higher orders on the $K$ factor is negligible.
The size of the shift due to the $K$ factor is two to three times as large in
the case of transversity with respect to helicity and typically reaches values
of the order of 10\% for these energies. It should be noted, however, that
there is an automatic limitation of the $K$-factor difference (\eg, with
respect to large values of $\alpha_s$ for low energies) owing to the presence
of the $\pi^2$ terms; when the $K$-factor difference between numerator and
denominator becomes large in absolute terms so too do the overall $K$ factors
themselves, which to some extent cancels or dilutes the effect. Very similar
corrections to those presented are obtained for energies corresponding to the
proposed experiments at GSI. Repeating the calculations for various
centre-of-mass energies, one finds that the corrections are maximal at
$\sqrt{s}\approx40$\,GeV, where for the transversity case they reach as much
as~15\%.

\subsection{Cross-checks}

A couple of simple cross-checks may be made on the influence of the scalar
vertex itself. First of all, there is now the possibility of a purely Higgs,
unpolarised \ac{DIS} process (\ie, in which both vertices are scalar). The
contribution of the real diagrams is
\begin{equation}
  \widetilde{C}^{\text{DIS-R},S}(z) =
  \widetilde{C}^\text{DIS-R}(z) - 2 - 3z ,
\end{equation}
which, combined with the virtual corrections already discussed, gives
\begin{equation}
  C^{\text{DIS},S}(z) =
  C^\text{DIS}(z)
  - 2 - 3z .
  \label{eq:combined-Higgs-DIS}
\end{equation}
The new coefficient in Eq.~\eqref{eq:combined-Higgs-DIS} could be used in place
of the usual unpolarised correction to \emph{define} the parton distributions.

Secondly, there is also similarly a possible purely scalar \ac{DY}-like
process; the \ac{NLO} correction to the unpolarised cross-section in this case
is found to be
\begin{equation}
  C^{\text{DY},S}(z) = C^\text{DY}\!(z) + \CF \, 2(1-z) .
\end{equation}
Moreover, Hikasa's theorem is avoided here owing to the presence of the scalar
vertices and a transverse-spin asymmetry is present even after integrating over
the lepton-pair azimuthal angle. The \ac{NLO} correction to the scalar
transversity asymmetry is
\begin{align}
  C_T^{\text{DY},S}(z) &=
  C^\text{DY}\!(z)
  \nonumber
\\[1ex]
  & \hspace{1.5em} \null +
  \CF \, (1-z) [ 4 - 4 \ln(1-z) + 2\ln z ] .
\end{align}

In none of the above cases does the scalar vertex introduce large correction
differences with respect to the vector. Indeed, in the last case of a purely
scalar \ac{DY} process (both spin averaged and transversely polarised), the
only differences are residues of the difference in the \ac{LO} splitting
functions, as indicated by the form and the overall factor $(1-z)$.

\section{Conclusions}

In order to appraise the real nature of the \ac{DY} $K$ factor in the case of
transversity asymmetries, we have examined \gedanken processes involving scalar
vertices. This allows a natural \ac{DIS} definition for the partonic densities
$\DT{q}(x)$. Such a definition allows a direct connection with model estimates
based on knowledge of parton densities derived essential from precisely
\ac{DIS}. Typical examples might be models in which at some low $Q^2$ scale
transversity- and helicity-weighted densities are naturally equal or others in
which the Soffer bound \cite{Soffer:1995ww} is found to be saturated, again at
some low scale. In all such cases the spin-averaged and helicity-weighted
densities used to set the starting point are obviously and naturally taken
directly from \ac{DIS}.

The results presented here provide a measure of the reliability of model
predictions, without, of course, representing a rigorous estimate, in as much
as the reference processes are partially fictitious and in any case are not
precisely those normally adopted. However, we have seen that in general the
corrections are not excessively large although they may be significantly larger
than in the helicity case. Moreover, comparison of the Mellin moments indicates
that in kinematical configurations in which low $z=\tau/x_1x_2$ dominates there
could be very important corrections. On the other hand, many of the differences
between \ac{NLO} coefficients vanish numerically for $z\to1$ and so safe
kinematical configurations certainly exist.

In closing then, although apparently fairly well under control, the question of
\ac{NLO} perturbative corrections in the case of transversely polarised \ac{DY}
processes clearly deserves more study, in particular, where the kinematics
might be such experimentally as to favour the dangerous low-$z$ region.

\section*{Acknowledgments}

Loop calculations were performed with the aid of FORM version 3.1
\cite{Vermaseren:2000nd}. The author would like to thank Johannes Bl\"{u}mlein for
very useful conversations in the past concerning the calculation of
transversity anomalous dimensions via photon--scalar interference diagrams.

\appendix

\section{Plus-regularised distributions}
\label{sec:Plus-reg}

The so-called ``plus-regularised'' distributions are defined via integrals with a
smooth test function $f(y)$:
\begin{multline}
  \int_x^1 \D{y} f(y) \left[\frac{g(y)}{1-y}\right]_+
  \equiv
  \int_x^1 \D{y} \left[\frac{f(y)-f(1)}{1-y}\right] \, g(y)
\\
  - f(1)\int_0^x \D{y} \frac{g(y)}{1-y}
  ,
\end{multline}
where $g(y)$ is well-behaved as $y\to1$.

\section{Modified minimal subtraction scheme: implementation}
\label{sec:Msbar}

The \ac{MS} scheme is defined, in conjunction with \ac{DR}, as the removal of
all simple poles in $1/\epsilon$ (double and higher poles due to \ac{IR}
divergences are cancelled automatically between real and virtual
contributions). However, common residual finite contributions are always left
due to the appearance of the factor $(4\pi)^\epsilon\Gamma(\epsilon)$.
Expanding to $\Order{\epsilon}$, one obtains
\begin{align}
  (4\pi)^\epsilon \Gamma(\epsilon)
  &\simeq
  \Gamma(1+\epsilon) \, \big[ 1 + \epsilon\ln(4\pi) \big] \, \frac1\epsilon
  \nonumber
\\
  &\simeq
  \frac1\epsilon + \ln(4\pi) - \gamma_\text{E}.
\end{align}
The \ac{MMS} scheme then augments \ac{MS} by subtracting the two
$\epsilon$-independent terms above.

Thus, \ac{MMS} may be implemented to $\Order{\alpha_s}$ by defining the Feynman
virtual momentum-integral measure $[\D[n]{k}]$ to include a factor
$1/\Gamma(1+\epsilon)$ and by removing a factor $(4\pi)^\epsilon$. In other
words, the plain \ac{MS} definition may be substituted with the following:
\begin{equation}
  [\D[n]{k}] \equiv
  \frac1{(4\pi)^\epsilon\Gamma(1+\epsilon)} \, \int \frac{\D[n]{k}}{(2\pi)^n}.
\end{equation}
Consequently, the definition of the phase-space integral for a final two-body
state (as in $\gamma{q}\to{qg}$) must be modified analogously to
\begin{equation}
  \text{PS}_2 \equiv
  \frac1{8\pi} \, s^{-\epsilon} \int_0^1 \D{y} \big[y(1-y)\big]^{-\epsilon},
\end{equation}
where $y=\frac12(1+\cos\theta)$ and $\theta$ is the partonic \ac{CM} scattering
angle (in this case between the incoming $q$ and outgoing $g$).

%——————————————————————————————————————————————————————————————————————————————%
% BibTeX bibliography
\bibliography{pigrostr,pigrotmp,pigropgr,pigrodbf,pigroxrf}
%——————————————————————————————————————————————————————————————————————————————%
\end{fmffile}
%——————————————————————————————————————————————————————————————————————————————%
\end{document}